\input ./style/arxiv-general.cfg
\documentclass[aoas,MSNbibl,nameyear,dvips]{arximspdf}
\makeatletter
   \@ifpackageloaded{graphicx}{}{\usepackage{graphicx}}
\makeatother

\doi{10.1214/15-AOAS817}
\volume{9}
\issue{2}
\pubyear{2015}
\firstpage{621}
\lastpage{639}
\docsubty{FLA}

\makeatletter
\def\mid{|}
\def\cal{\mathcal}

\newcommand{\eqref}[1]{(\ref{#1})}
\def\btheta{\bolds{\theta}}
\def\bmu{\bolds{\mu}}

\newcommand{\bx}{\mathbf{x}}
\newcommand{\bN}{\mathbf{N}}
\newcommand{\bZ}{\mathbf{Z}}
\newcommand{\bw}{\mathbf{w}}
\newcommand{\ba}{\mathbf{a}}

\newcommand{\Cs}{C} 
\newcommand{\Cstar}{C^\star}
\newcommand{\true}{{\mathrm{TRUE}}}
\renewcommand{\th}{\theta}
\newcommand{\eps}{\varepsilon}
\newcommand{\bth}{\bolds{\th}}
\newcommand{\Ct}{\widetilde{C}}
\newcommand{\bxt}{\widetilde{\bx}}

\newcommand{\DD}{{\cal D}}
\newcommand{\Ga}{\operatorname{Gamma}}
\newcommand{\Dir}{\operatorname{Dir}}
\newcommand{\Be}{\operatorname{Be}}
\newcommand{\Binom}{\operatorname{Bin}}

\newcommand{\iid}{\stackrel{\mathrm{i.i.d.}}{\sim}}
\newcommand{\indep}{\stackrel{\mathrm{indep}}{\sim}}

\makeatother

\begin{document}
\begin{frontmatter}

\title{A Bayesian feature allocation model for~tumor~heterogeneity}
\runtitle{Bayesian model for tumor heterogeneity}

\begin{aug}
\author[A]{\fnms{Juhee}~\snm{Lee}\thanksref{M1}}, 
\author[B]{\fnms{Peter}~\snm{M\"uller}\thanksref{M2,T1}\ead[label=e2]{pmueller@math.utexas.edu}},
\author[C]{\fnms{Kamalakar}~\snm{Gulukota}\thanksref{M3}} 
\and
\author[D]{\fnms{Yuan}~\snm{Ji}\corref{}\thanksref{M3,M4,T1}\ead[label=e4]{jiyuan@uchciago.edu}}
\thankstext{T1}{Supported in part by NIH R01 CA132897.}
\runauthor{Lee, M\"uller, Gulukota and Ji}
\affiliation{University of California Santa Cruz\thanksmark{M1}, University of Texas, Austin\thanksmark{M2},
NorthShore University HealthSystem\thanksmark{M3} and University of Chicago\thanksmark{M4}}
\address[A]{J. Lee\\
Applied Mathematics \& Statistics\\
Baskin School of Engineering\\
University of California\\
1156 High Street\\
Mail Stop SOE2\\
Santa Cruz, California 95064\\
USA}
\address[B]{P. M\"uller\\
Department of Mathematics\\
University of Texas, Austin\\
1 University Station, C1200\\
Austin, Texas 78712\\
USA\\
\printead{e2}\hspace*{35pt}}
\address[C]{K. Gulukota\\
NorthShore University HealthSystem\\
1001 University Place\\
Evanston, Illinois 60201\\
USA}
\address[D]{Y. Ji\\
NorthShore University HealthSystem\\
1001 University Place\\
Evanston, Illinois 60201\\
USA\\
and\\
Department of Public Health Sciences\\
University of Chicago\\
5841 South Maryland Ave MC2000\\
Chicago, Illinois 60637\\
USA\\
\printead{e4}}
\end{aug}

%
\received{\smonth{7} \syear{2014}}
%
\revised{\smonth{1} \syear{2015}}

%
\begin{abstract}
We develop a feature allocation model for inference on genetic tumor
variation using next-generation sequencing data. Specifically, we
record single nucleotide variants (SNVs) based on short reads mapped
to human reference genome and characterize tumor heterogeneity by
latent haplotypes defined as a scaffold of SNVs on the same homologous
genome. 
For multiple samples from a single tumor, assuming that each sample is
composed of some sample-specific
proportions of these haplotypes, 
we then fit the observed
variant allele fractions of SNVs for each sample and
estimate the proportions of haplotypes.
Varying
proportions of haplotypes across samples is evidence of tumor
heterogeneity since it implies varying composition of cell
subpopulations.
Taking a Bayesian perspective, we proceed with a prior probability
model for all relevant unknown quantities, including, in particular,
a prior probability model on the binary indicators that characterize
the latent haplotypes.
Such prior models are known as feature allocation models. Specifically,
we define a simplified version of the Indian buffet process, one of
the most traditional feature allocation models. The proposed model
allows overlapping clustering of SNVs in
defining latent haplotypes, which reflects the evolutionary process of
subclonal expansion in tumor samples.
\end{abstract}

%
\end{frontmatter}

\section{Introduction}
\label{sec:Intro_intra}

We propose a feature allocation model
[Broderick, Jordan and Pitman (\citeyear{Tama:Jord:Pitm:sts:2013,Tama:Pitm:Jord:2013})] to describe tumor
heterogeneity using next-generation sequencing (NGS) data.
We use a variation of the Indian buffet process (IBP)
[\citet{griffiths2005infinite,teh2007stick}].
The feature allocation in our model is latent. That is, the features
are not directly observed. We record point mutations as single nucleotide
variants (SNVs), each of which is defined as a DNA
locus that possesses a variant sequence from that on the
reference human genome. We use the feature allocation model
to describe unobserved haplotypes, defined as a collection of
single nucleotide variants (SNVs) scaffolded on a homologous genome. In a
tumor sample, having more than two haplotypes is evidence of
heterogeneous cell subpopulations with a distinct genome.
This is the case because humans are diploid and
we would therefore only observe up to
two haplotypes if all cells in a tumor sample were
genetically homogeneous. In the proposed application of feature
allocation models to
inference for tumor heterogeneity,
the haplotypes are the
features and the SNVs are the experimental units that select the
features. 
The number of features is unknown. Each
tumor sample is composed of an unknown proportion of each of these
haplotypes. The top level sampling model for the observed SNV counts
is then defined as binomial sampling with a proportion for each SNV
that is implied by this composition. In summary, we solve a
deconvolution problem to explain the observed SNV frequencies for each
sample by compositions of latent haplotypes.

Heterogeneity in cancer tissue has been hypothesized over the past
few decades [\citet{oldhetero}] and has been demonstrated elegantly
using NGS technology over the past few years
[\citet{newhetero}]. Genetic variation in a tumor occurs due to
evolutionary processes that drive tumor progression.
Specifically, tumors include
distinct clonal subpopulations of cells that arise stochastically by a
sequence of randomly acquired
mutations. Substantial genetic heterogeneity between tumors
(inter-tumor) or
within a tumor (intra-tumor) can be explained by differences in clonal
subpopulations and varying proportions of those
subpopulations [\citet
{Marusyk:2010:intra,Russnes:hetero:2011,CLL:2013:cell}]. For example,
\citet{navin2010inferring}
reported clonal genomic heterogeneity in breast cancers.


Data derived from NGS experiments include SNVs,
small indels and copy number variations
[\citet{wheeler2008complete,ng2010whole}].
Many researchers use SNV data to investigate genes and
genomic regions related to cancer phenotypes [\citet
{erichsen2004snps,engle2006using}].
In this paper, we utilize whole-genome sequencing data measuring
variant allele fractions (VAFs) at SNVs to understand tumor
heterogeneity by proposing inference on
how haplotypes may be distributed within a tumor.

\begin{figure}[b]
\centering
\begin{tabular}{@{}c@{}}

\includegraphics{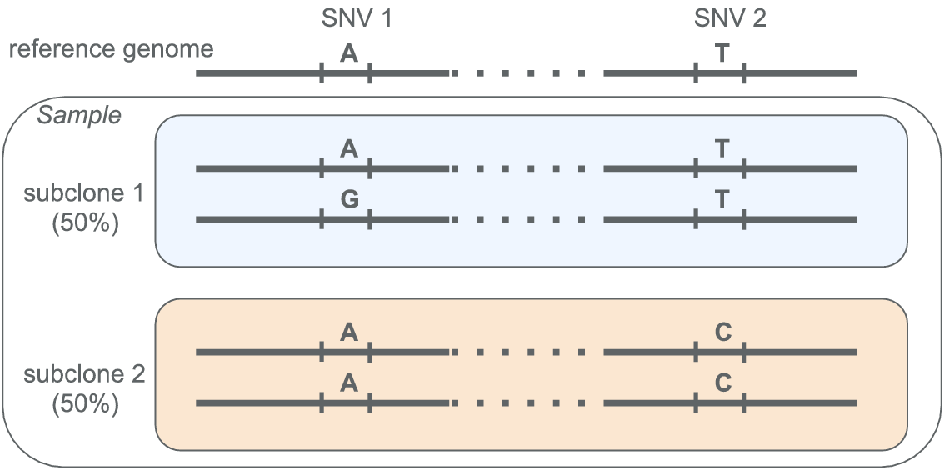}
\\
\footnotesize{\textup{(a)} Multiple haplotypes as evidence of
a heterogeneous tumor}\\[3pt]

\includegraphics{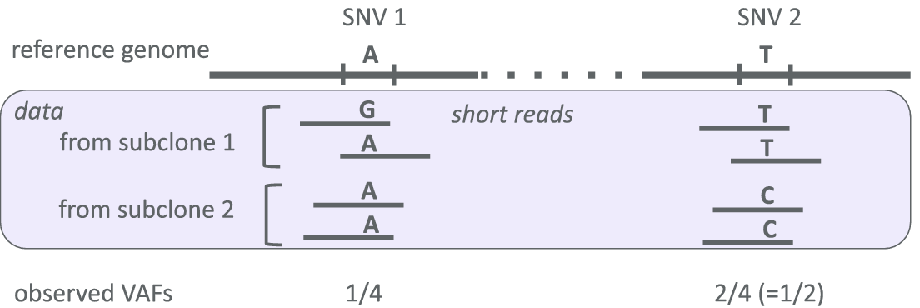}
\\
\footnotesize{(b) Hypothetical short reads data}
\end{tabular}
\caption{A hypothetical example explaining how NGS data can be
used to infer heterogeneous tumor samples. \textup{(a)} shows that there
are two subclones (cell subpopulations) in the tumor sample with
different haplotypes consisting of two SNVs: For subclone 1, there
are two haplotypes, AT and GT. For subclone 2, there is only
one haplotype, AC. Thus, there are a total of three haplotypes in
the tumor sample, implying heterogeneous cell populations since a
population of
homogeneous cells would only support up to two haplotypes. Here
sequence G for SNV 1 and sequence C for
SNV 2 are mutations.
\textup{(b)} shows hypothetical short reads for this sample if it is
sequenced, assuming that the proportions of the two subclones
are equal. The short reads counts
are summarized as observed VAFs, which
are used for our statistical inference.}
\label{fig:data_1}
\end{figure}

In an NGS experiment, millions of short DNA reads
are generated and are then aligned to the reference genome. At certain positions
of the genome, 
some or all of the mapped reads will show a sequence different from
what is in the reference genome. At each genomic locus, the proportion
of short reads
bearing a variant sequence is called the VAF. If the VAF at a locus
is nonzero, an SNV may be ``called'' at that locus, based on
statistical inference [\citet{li2009sequence}]. The raw
experimental data comprises the total number of reads ($N$)
that are mapped to the locus and the number of those reads ($n$)
indicating a variation from the reference sequence. Then VAF $=
n/N$.
If a tumor sample is homogeneous, that is, having a single clone, the VAF
values of all the SNVs should be close to 0, 0.5 or 1, reflecting the
three possible homozygous and heterozygous alleles (i.e., AA, AB,
BB) at any SNV. Different VAF values imply
heterogeneity of the cellular genome in the
tumor sample (see Figure~\ref{fig:data_1} for an example).
We propose to study inference to deconvolute the VAFs from
multiple SNVs and back out the latent haplotypes.

%



\begin{figure}

\includegraphics{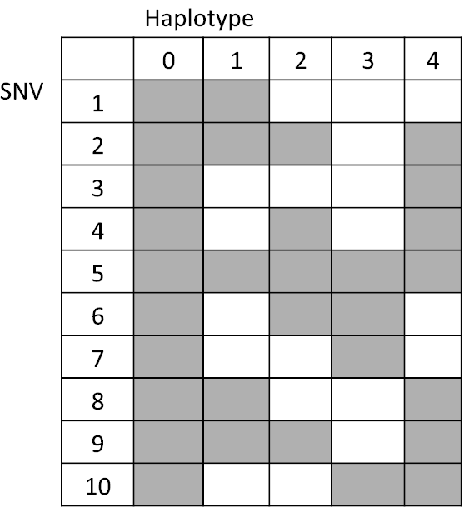}

\caption{An illustration of cell types (binary latent features) in
columns. A black/white block indicates a variant/reference DNA
sequence at the
corresponding SNV (row) for the haplotype (column).}
\label{fig:demo_Z}
\end{figure}

We propose a Bayesian feature allocation model to characterize
such cellular heterogeneity in a way that explains the observed NGS
data.
We construct a matrix of binary features
(equivalently, haplotypes) as shown in Figure~\ref{fig:demo_Z}.
In the figure, columns correspond to haplotypes and rows correspond to
SNVs. We define haplotype $c$ by a binary vector
$(z_{1c},\ldots,z_{Sc})$ of indicators of whether ($z_{sc}=1$) or not
($z_{sc}=0$) a variant sequence is observed at the SNV $s$. Here
we view SNV as a genetic locus on which either a variant or
reference DNA sequence could be observed.
Figure~\ref{fig:demo_Z}
illustrates the definition of five haplotypes ($C=4$, columns) with $S=10$
SNVs (rows).
In the figure, black (white) indicates $z_{sc}=1$ ($z_{sc}=0$).
For example, SNV 1 in Figure~\ref{fig:demo_Z} possesses a variant
sequence in the two haplotypes $c=0$ and $c=1$. On the other
hand, SNV 9 possesses variant sequences in four haplotypes:
$c=0, 1, 2$ and $4$.
A prior probability model on such a binary matrix $\bZ=[z_{sc}]$ is
known as a feature allocation model. Here, we assume that $C$ is
unknown and place a prior on $C$.

%

Assuming that samples are composed of different proportions of
$C$ haplotypes, we aim to fit the observed VAFs of
the SNVs to infer these proportions.
For example, we may observe that one sample is dominated by
haplotypes $1$ and $4$, while another is dominated by haplotypes $2$ and
$3$. If the samples are from the same tumor, the differences in
haplotypic 
compositions are evidence of \textit{intra-tumor}
heterogeneity; on the other hand, differences in samples from
different tumors imply \textit{inter-tumor} heterogeneity. Therefore,
the proposed inference provides a unified framework to address
inference for both biological concepts.
Importantly, the characterization of haplotypes is based on
selected SNVs only. Otherwise inference for 
heterogeneity between tumors in different patients
would not be biologically meaningful, as
cellular genomes and haplotypes are not expected to be shared across
patients. However, for tumors in the same class of disease,
SNVs in local disease-related genomic regions
may be common to all or some of the tumors, thereby allowing for the
proposed inference.



There are currently few approaches that address the problem of tumor
heterogeneity. \citet{su2012purityest} and \citet{PurBayes:2013}
recognized that a tumor sample is a mixture of normal cells and tumor
cells, and developed a method to estimate tumor purity levels for
paired tumor-normal tissue samples using DNA sequencing data. None of
the two methods considers more than two samples or unpaired
samples. 
PurBayes [\citet{PurBayes:2013}] accounts for
intra-tumor heterogeneity, but it does not provide inference on the
subpopulation configurations as inference on the latent matrix $\bZ$
under the proposed model. PyClone [\citet{roth2014pyclone}], a
recently published
method, proposes inference to cluster SNVs with different VAFs. An
underlying assumption of PyClone is that
SNVs can be arranged in clusters that inform about subclones.
A key component of PyClone is the use of
clustering models such as the Dirichlet process for inference
on these clusters.
While such clusters are informative about heterogeneity, inference
that is provided by PyClone is not meant to identify subclones or
haplotypes.
The primary aim of PyClone is inference on mutation clusters, defined
as a group of SNVs with similar variant allele fractions.

In contrast, our proposed feature
allocation model explicitly models the haplotypic genomes of
subclones, allowing overlapping SNVs shared between different
subclones. We do not use nonoverlapping SNV clusters as the building
block for subclones. That is, instead of first estimating the SNV
clusters and then constructing subclones based on clusters, we
directly infer the subclonal structure based on haplotypes. We show in
later examples the distinction between PyClone and our proposed
method.

The remainder of the paper is organized as follows:
Section~\ref{sec:Model_Sel} describes the proposed Bayesian feature
allocation model and a model selection criterion to select the number
of subclones. Section~\ref{sec:Simulation} describes simulation
studies. Sections~\ref{sec:Example} and~\ref{sec:lung_cancer} report
data analyses of in-house data sets to illustrate inter-tumor
heterogeneity and intra-tumor heterogeneity, respectively. The last
section concludes with a final discussion.

%
\begin{figure}[t]

\includegraphics{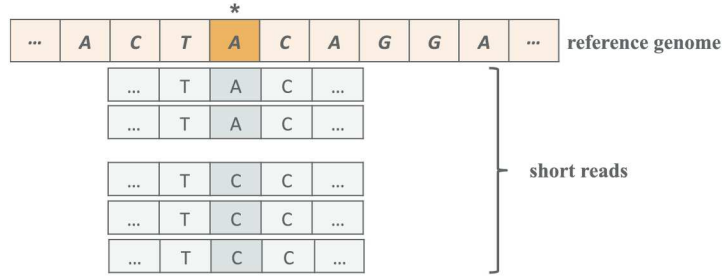}

\caption{An illustration of the Binomial model.
The illustration shows that 5 short reads are
mapped to a position marked with * and among
them three reads indicate variation at the position,
that is, $N_{st}=5$ and $n_{st}=3$.}
\label{fig:demo_bin}
\end{figure}
%

\section{Probability model}
\label{sec:Model_Sel}
\subsection{Sampling model}
\label{sec:Models}
Let $\mathbf{n}$ denote an $S \times T$ matrix of count data
from an
NGS genome sequencing experiment, with $n_{st}$ denoting the
number of reads that bear a variant sequence at the location of SNV $s$
for tissue sample $t$, $s=1,
\ldots, S$ and $t=1, \ldots, T$. 
We assume a binomial sampling
model. Let $N_{st}$ denote the total number of reads in sample $t$ that
are mapped to the genomic location of SNV $s$.
We assume
%
\begin{equation}
n_{st} \indep\Binom(N_{st}, p_{st}). \label{eq:like}
\end{equation}
In Figure~\ref{fig:demo_bin}, $n_{st}=3$ and $N_{st}=5$. We do not
model $N_{st}$, that is, we treat $N_{st}$ as fixed, and only
consider a sampling model for $n_{st}$ conditional on $N_{st}$
(modeling $N_{st}$ would not contribute any information
on tumor heterogeneity based on SNVs).
Conditional on $N_{st}$, the observed counts $n_{st}$ are independent
across $s$ and $t$. The model in \eqref{eq:like} is illustrated in
Figure~\ref{fig:demo_bin}.

\subsection{Prior}
We build a prior probability model for $p_{st}$ in two steps, using
the notion that each sample is composed of a mixture of different
haplotypes. And each haplotype, in turn, is characterized by the
haplotypes consisting of the SNVs. Let $w_{tc}$ denote the
proportion 
of haplotype
$c$ in sample $t$ and let $z_{sc} \in\{0,1\}$ denote a latent
indicator of the event that SNV $s$ bears a variant sequence for
haplotype $c$. Note that $z_{sc} =1$ corresponds to a black block in
Figure~\ref{fig:demo_Z}. 
Then $p_{st}$ is
written as a sum over $C$ latent haplotypes
%
\begin{equation}
p_{st} = w_{t0}p_0 + \sum
_{c=1}^C w_{tc} z_{sc} \equiv
\varepsilon_{t0} + \sum_{c=1}^C
w_{tc} z_{sc}. \label{eq:model_p}
\end{equation}
%
The construction of the haplotypes, including the number of haplotypes, $C$,
and the indicators $z_{sc}$ are latent. The key term, $\sum_{c=1}^C
w_{tc}z_{tc}$, indirectly infers haplotypes 
by explaining $p_{st}$ as arising from
sample $t$ being composed of a mix of hypothetical haplotypes which do
($z_{sc}=1$) or do not ($z_{sc}=0$) include a mutation for SNV~$s$.
The indicators $z_{sc}$ are collected in a $(S \times C)$ binary matrix
$\mathbf{Z}$. The number of latent haplotypes, $C$, is unknown. Conditional on
$C$, the binary matrix $\mathbf{Z}$
describes $C$ latent tumor haplotypes present 
in the observed
samples. Joint inference on $C$, $Z$ and $\bw_t$ explains tumor
heterogeneity. 

In addition, we introduce a background haplotype, labeled as
haplotype $c=0$, which includes all SNVs. The background haplotype
accounts for experimental noise and haplotypes that appear with
negligible abundance. Specifically, $\varepsilon_{t0}=w_{t0}p_0$
in \eqref{eq:model_p} relates to this background haplotype, with $p_0$
being the relative frequency of observing a mutation at an SNV
due to noise and artifact (we assume equal frequency for all SNVs) and
$w_{t0}$ being the proportion in sample $t$. The prior on $w_{t0}$ is
defined later.
For $p_0$, we assume $p_0 \sim\Be(a_{00}, b_{00})$ with $a_{00}
\ll b_{00}$ to inform a small $p_0$ value a priori.

We start the prior construction with a prior for the number of
haplotypes, $C$. We consider a geometric distribution, $C \sim
\mbox{Geometric}(r)$ where $\mathrm{E}(C)=1/r$. Conditional on $C$, we
use a feature allocation model for a binary matrix $\bZ$. 
We first define the model for any given $C$ and
start with feature-specific selection probabilities,
%
\begin{equation}
\mu_c \mid C \iid\Be(\alpha/\Cs, 1) \label{eq:p_mu}.
\end{equation}
Let $\bmu=(\mu_1, \ldots, \mu_{\Cs})$.
The selection probabilities are used to define
$p(\bZ\mid\bmu, C)$ as
%
\begin{equation}
\mathrm{p}(\bZ\mid\bmu, C) = \prod_{s=1}^S
\prod_{c=1}^{\Cs}\mu _{c}^{z_{sc}}(1-
\mu_{c})^{(1-z_{sc})} = \prod_{c=1}^{\Cs}
\mu_c^{m_c}(1-\mu_c)^{S-m_c},
\label{eq:cond_Z}
\end{equation}
where $m_c=\sum_{s=1}^S z_{sc}$ is the number of SNVs bearing
variant sequences for 
haplotype $c$. 
A limit of the model, as $C \rightarrow\infty$,
becomes a constructive definition of the Indian buffet process (IBP)
[\citet{griffiths2005infinite,teh2007stick}].
The model is symmetric with respect to arbitrary indexing of
the SNVs, simply because of the symmetry in \eqref{eq:cond_Z} and
\eqref{eq:p_mu}. Note that $m_c=0$ is possible with positive prior probability.

Next, we consider a prior distribution for abundances associated with
the haplotypes defined by $\bZ$. The haplotypes are common for
all tumor samples, but the relative weights in the composition \eqref
{eq:model_p} are
different across tissue samples.
We assume Dirichlet priors for the relative weights $w_{tc}$,
defined as follows.
Let $\theta_{tc}$ denote an (unscaled) abundance level of haplotype
$c$ in
tissue sample $t$.
We assume $\theta_{tc} \mid C \iid\Ga(a, 1)$ for $c=1, \ldots, \Cs
$ and
$\theta_{t0} \iid\Ga(a_0, 1)$. We then define
\[
w_{tc}= \theta_{tc}\Big/\sum_{c^\prime=0}^{\Cs}
\theta_{tc^\prime} %
\]
as the relative weight of haplotype $c$ in sample $t$.
This is equivalent to $\bw_t \mid C \iid\Dir(a_0, a, \ldots, a)$ for
$t=1, \ldots,
T$, where $\bw_t=(w_{t0}, w_{t1}, \ldots, w_{t\Cs})$.

Recall the binomial sampling likelihood \eqref{eq:like} with success
probability, $p_{st}$. Given $C$, $\bZ$ and $\bw$, we define $p_{st}$
in \eqref{eq:model_p}. In words, $p_{st}$ is determined by $C$, $\bZ$
and $\bw_t$ with the earlier describing the latent haplotypes
and the latter specifying the relative abundance
of each haplotype in sample $t$.

\subsection{Posterior simulation}
\label{sec:post_sim}
Let $\bx= (\bZ, \bth, p_0)$, where $\bth= \{\theta_{tc}\}$.
Markov chain Monte Carlo (MCMC) posterior simulation proceeds by
sequentially drawing random numbers for the parameters in $\bx$. Given
$C$, such MCMC simulation is straightforward. 
Specifically, Gibbs sampling transition probabilities
are used to update $z_{sc}$, and Metropolis--Hastings transition
probabilities are used to update $\btheta$ and $p_0$. It is possible
to improve the mixing of the Markov chain by updating all columns in
row $s$ of the matrix $\bZ$ jointly by means of a Metropolis--Hastings
transition probability that proposes changes in the entire row vector
$\mathbf{z}_s$.

The construction of transition probabilities that involve a
change of $C$ is more challenging, since
the dimension of $\bZ$ and $\bth$ changes as $C$ varies.
We use a reversible jump (RJ) MCMC algorithm
for posterior simulation [\citet{green1995reversible}].
We first define a proposal distribution
$q(C,\Ct)$ for $C$, and then introduce a proposal distribution $q(\bxt
\mid\Ct)$ for $\bx$ conditional on the proposed $\Ct$. The latter
potentially involves a change in dimension of the parameter vector. We
found that high posterior correlation of $\bZ$ and
$\bw$ conditional on $C$ greatly complicated
the construction of a practicable RJ scheme.
To overcome this, we use an approach similar to
\citet{casella2006objective}.
We split the data into a minimal training set $(\mathbf{n}',\bN')$
with $n_{st}^\prime=b_{st} n_{st}$, $N'_{st}=b_{st} N_{st}$,
and a test data set, $(\mathbf{n}^{\prime\prime},\bN
^{\prime\prime})$ with
$n^{\prime\prime}_{st}=(1-b_{st})
n_{st}$ etc.
In the implementation we use $b_{st}$ generated from $\Be(25, 975)$.
Let $p_1(\bx\mid C) = p(\bx\mid\mathbf{n}^{\prime},C)$
denote the
posterior distribution under $C$ using the training sample.
We use $p_1$ in two instances.
First, we replace the original prior $p(\bx\mid C)$ by $p_1(\bx\mid C)$
and, second,\vspace*{1pt} we also use it as proposal distribution $q(\bxt\mid\Ct) =
p_1(\bxt\mid\Ct)$.
The test data is then used to evaluate the acceptance probability.
The strategy can be characterized as model comparison by
fractional Bayes factors [\citet{OHagan:95}] and is related to a similar
approach proposed in \citet{casella2006objective} for model comparison
with intrinsic Bayes factors.
Both are originally proposed for model comparison with noninformative
priors, but can be modified to facilitate MCMC across models as we need
it here.

\begin{figure}[b]
\centering
\begin{tabular}{@{}cc@{}}

\includegraphics{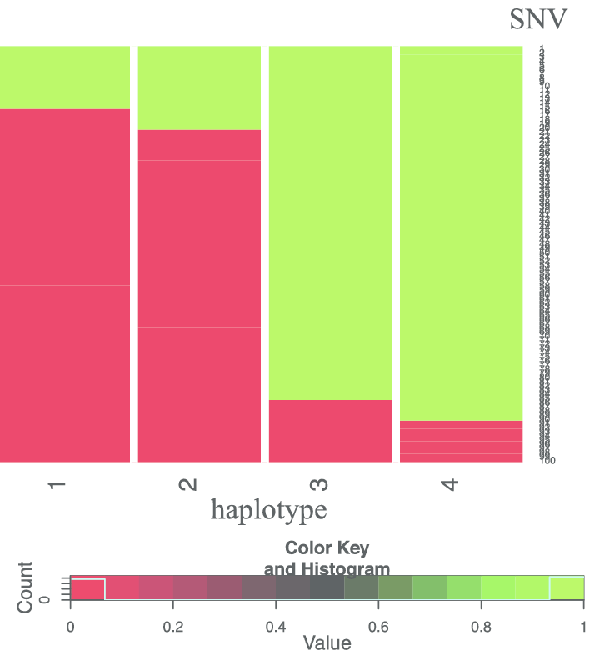}  & \includegraphics{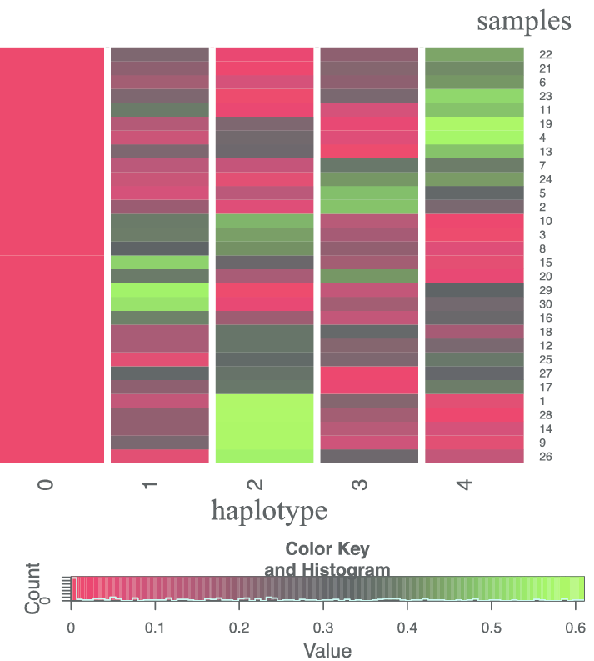}\\
\footnotesize{(a) $\bZ^\true$} & \footnotesize{(b) $\bw^\true$}
\end{tabular}
\caption{Heatmaps of $\bZ^\true$ and $\bw^\true$ in the
simulation.}
\label{fig:Sim_tr}
\end{figure}

We summarize the joint posterior distribution, $p(C, \bZ, \bw, p_0
\mid\mathbf{n})$, by factorizing it as
$p(C \mid\mathbf{n})p(\bZ\mid\mathbf{n}, C)
p(\bw,p_0 \mid\mathbf{n}, C, \bZ)$.
Based on the available posterior Monte Carlo sample, we
(approximately) evaluate the marginal posterior distribution for $C$ and
determine the maximum a posteriori (MAP) estimate $C^\star$.
We then estimate $\bZ$ conditional on $C^\star$ as follows:
For any two matrices, $\bZ$ and $\bZ'$, $1 \le c, c^\prime\le
C^\star
$, let
$\DD_{cc'}(\bZ,\bZ')=\sum_{s=1}^S |z_{sc} - z^\prime_{sc^\prime}|$.
We then define a distance 
$
d(\bZ, \bZ^\prime) = \min\sum_{c=1}^{C^\star} \DD_{c,\pi_c}(\bZ
,\bZ')$,
where $\pi_c$ is a permutation of $\{1, \ldots, C^\star\}$ and the minimum
is over all possible permutations.
A posterior
point estimate for $\bZ$ is defined as
\[
\bZ^\star_C = \operatorname{arg} \min_{\bZ^\prime} \int d\bigl(
\bZ, \bZ^\prime\bigr) \,dp\bigl(\bZ \mid\mathbf{n},
C^\star\bigr) \approx \operatorname{arg} \min_{\bZ^\prime} \frac{1}{L}
\sum_{\ell=1}^L d\bigl(\bZ^{(\ell)},
\bZ^\prime\bigr), %
\]
for a posterior Monte Carlo sample, $\{\bZ^{(\ell)}, \ell=1, \ldots
, L\}$.
Finally, we report posterior point estimates $\bw^\star$ and
$p_0^\star$
for $\bw$ and $p_0$ conditional on $C^\star$ and $\bZ^\star_C$.

\section{Simulation}
\label{sec:Simulation}
We validated the proposed model in a simulation study.
We simulated a set of
$S=100$ SNVs with $T=30$ samples. In the simulation truth,
we assumed four latent haplotypes ($C^{\true}=4$) as well as a
background haplotype ($c=0$) with all SNVs bearing variant
sequences. 
Haplotype $c=1$ has variant sequences for the first 15 SNV positions,
haplotype 2 for the first 20
SNV positions, haplotype 3 for the first 85 positions and haplotype 4
for the first 90 positions. In other words, SNVs 1--15 bear
variant sequences for all
four haplotypes, SNVs 16--20 for haplotypes 2--4, SNVs 21--85 for
haplotypes 3--4, SNVs 86--90 for haplotype~4 only and SNVs 91--100 for
none of the haplotypes, as shown in Figure~\ref{fig:Sim_tr}(a). The green
color in panel (a)
implies presence ($z_{sc}=1$) of a variant sequence at SNV
$s$ for haplotype $c$ and the red color shows absence
($z_{sc}=0$), for $c=1, \ldots, 4$ and $s=1, \ldots, 100$.
We then generated $\bw_{t}^\true$ as follows.\vspace*{1pt} 
We let $\ba^\true=(8, 6, 3, 1)$ and for each $t$ randomly
permuted $\ba^\true$. Let\vspace*{1pt} $\ba_\pi^\true$ denote a random permutation
of $\ba^\true$. We generated $\bw^\true\sim\Dir(0.2,
\ba_\pi^\true)$. That is, the first parameter of the Dirichlet prior
for the $(C^\true+1)$-dimensional weight vector was $0.2$, and the remaining
parameters were a permutation of $\ba^\true$. Using the assumed
$\bZ^\true$ and $\bw^\true$ and letting $p_0^\true=0.01$ and
$N_{st}=50$ for all $t$ and $s$, we generated $n_{st} \sim
\Binom(N_{st}, p^\true_{st})$ with $p^\true_{st}=p_{0}^\true
w^\true_{t0}+\sum_{c=1}^{\Cs}w^\true_{tc}z^\true_{sc}$. The
weights $\bw
^\true$
are shown in Figure~\ref{fig:Sim_tr}(b). Similar to the
heatmap of $\bZ^\true$, the green color in panel (b)
represents high abundance of a haplotype in a sample and the red color low abundance for $c=0, \ldots, 4$ and $t=1, \ldots, 30$.
For haplotype 0 the heatmap 
plots $w_{t0}p_0$. The samples in rows are rearranged for better display.

To fit the proposed model, we took $r=0.2$, $\alpha=3$, $a_0=0.5$,
$a=0.5$, $a_{00}=1$ and
$b_{00}=100$. For each value of $C$, we initialized $\bZ$ using the
observed sample proportions. We generated initial values for
$\btheta_{tc}$ and $p_0$ by prior draws. We generated $b_{st} \iid\Be
(25, 975)$ to construct the minimal training set and ran the MCMC
simulation over 25,000 iterations, discarding the first 10,000
iterations as initial burn-in.

%
\begin{figure}[b]
\centering
\begin{tabular}{@{}cc@{}}

\includegraphics{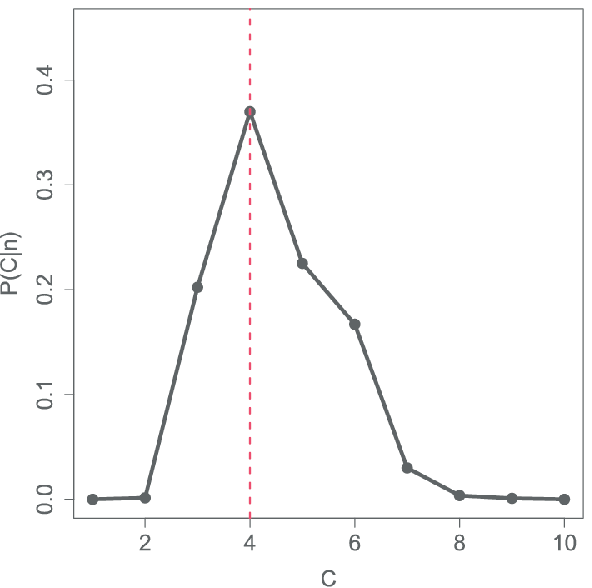}  & \includegraphics{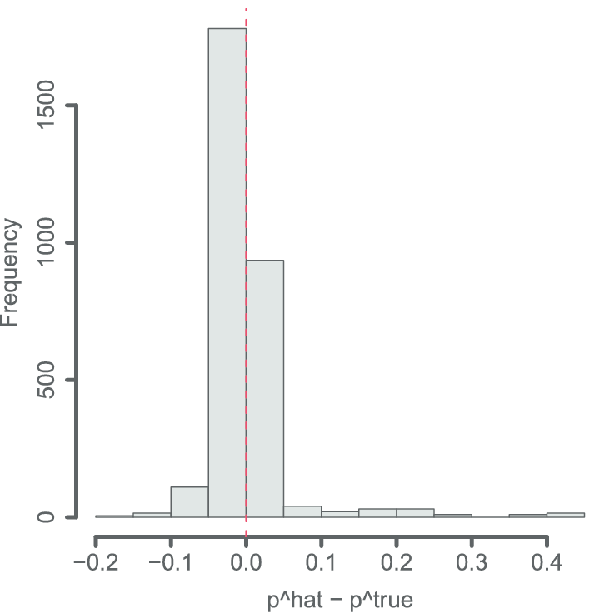}\\
\footnotesize{(a) Posterior distribution of $C$} & \footnotesize{(b)
Histogram of $\hat{p}_{st} - p_{st}^{\mathrm{TRUE}}$}\\[3pt]

\includegraphics{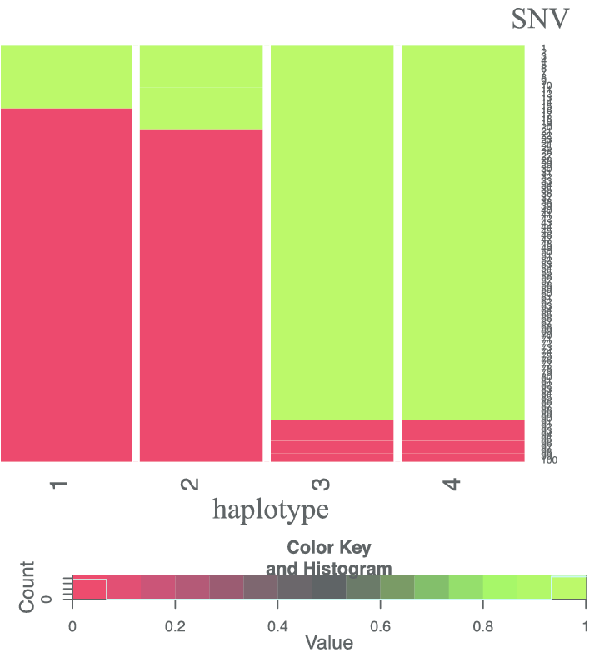}  & \includegraphics{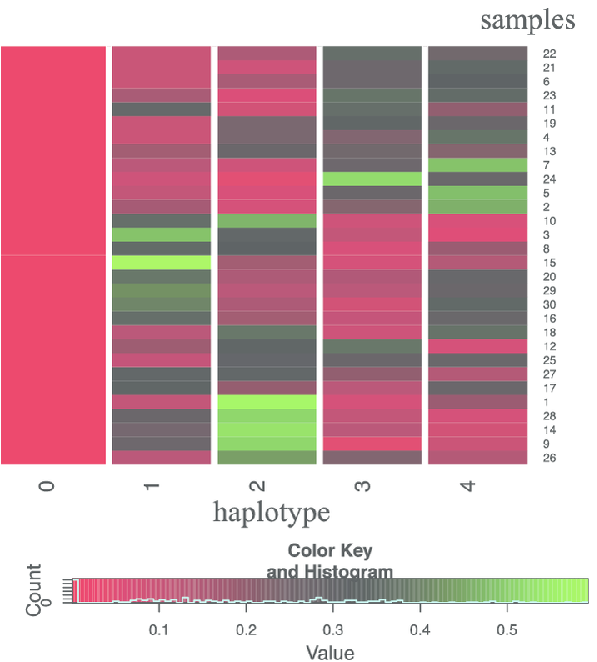}\\
\footnotesize{(c) Heatmap of $\bZ^\star_C$ with $C^\star=4$} &
\footnotesize{(d) Heatmap of $\bw^\star_C$ with $C^\star=4$ and
$\bZ
^\star_C$}
\end{tabular}
\caption{Posterior inference for the simulated data.}
\label{fig:Sim_result}
\end{figure}


Figure~\ref{fig:Sim_result}(a) reports the posterior distribution of $C$
in which the dashed vertical line represents the true value
$C^{\true}=4$. The posterior mode $\Cstar=4$ recovers the truth.
We then find the
posterior point estimates of $\bZ$, $\bw$ and $p_0$ conditional on
$C^\star$ as described in Section~\ref{sec:post_sim}. We compared $p^\true_{st}$
with the
posterior estimates
$\hat{p}_{st}=p^\star_0w_{t0}^\star+
\sum_{c=1}^{C^\star}w_{tc}^\star z_{sc}^\star$.\vspace*{1pt}
Figure~\ref{fig:Sim_result}(b) shows the histogram of the errors
$(\hat{p}_{st}-p^\true_{st})$. Fitting appears to be great as
$(\hat{p}_{st} - p_{st})$ centers at $0$. 
Figure~\ref{fig:Sim_result}(c) and (d) show
heatmaps for
$\bZ^\star_C$ and $\bw^\star_C$ (given $C^\star=4$).
The estimate $\bZ^\star_C$ in Figure~\ref{fig:Sim_result}(c) places SNV
86--90 into haplotypes 3 and\vspace*{1pt} 4. The latter are two identical haplotypes.
This may be because $w_{t3}^{\mathrm{TRUE}}$ for haplotype $c=3$ is small for
almost all the samples.
The weights for the two dominant subclones, $w^\star_{tc}$ for
$c=1, 2$, are close to the simulation truth, and
$w^\star_{tc}$ for $c=3,4$ are closer to the
average of $w_{tc}^{\true}$, $c=3, 4$.


\begin{figure}[b]
\centering
\begin{tabular}{@{}cc@{}}

\includegraphics{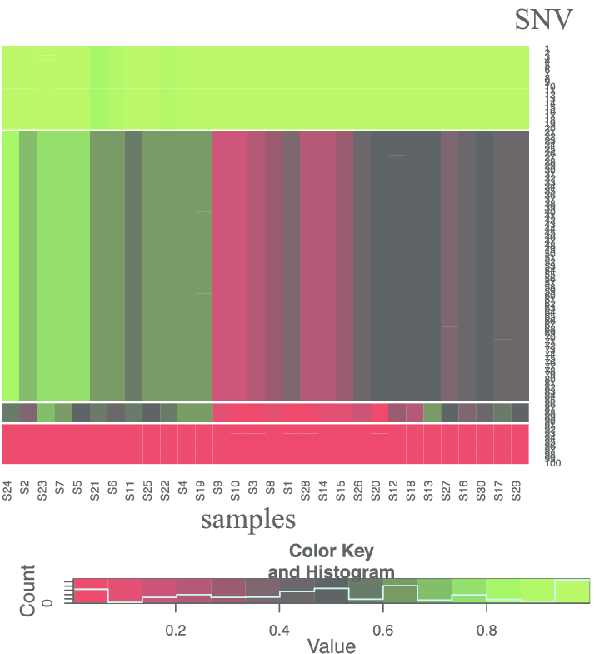}  & \includegraphics{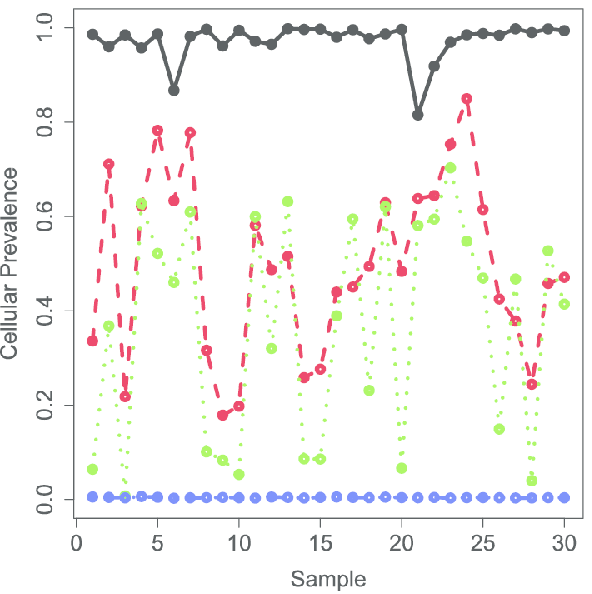}\\
\footnotesize{(a) Heatmap of cellular prevalence} & \footnotesize{(b)
Mean prevalence for each cluster}
\end{tabular}
\caption{Estimated cellular prevalence of four SNV clusters over
samples by PyClone for the simulated data.}
\label{fig:Sim_pyclone}
\end{figure}
%

For comparison, we implemented PyClone [\citet{roth2014pyclone}] with
the same simulated data.
We used the infinite beta-binomial mixture model in PyClone
assuming that the copy number at mutation positions is
known. Figure~\ref{fig:Sim_pyclone}(a) shows the estimated variant allelic
prevalence for each mutation \textit{for each sample} under PyClone.
Columns are samples and rows are SNVs.
The white horizontal lines separate the estimated SNV
clusters. 
PyClone identified four clusters of SNVs: cluster 1 with SNV 1--20,
cluster 2 with SNV 21--85, cluster 3 with SNV 86--90 in cluster 3,
and cluster 4 with SNV 91--100.

The estimated cluster 1 includes the SNVs that under the simulation truth
appear in all the true haplotypes or in true haplotypes 2--3;
cluster 2 includes the SNVs from true haplotypes 3--4;
cluster 3 includes SNVs from true haplotype 4;
and cluster 4 includes SNVs that appear in none of the true haplotypes.
Figure~\ref{fig:Sim_pyclone}(b) shows estimated mean
cellular prevalence of each cluster across the 30 samples. In summary,
the reconstruction under PyClone is reasonable, but
stops short of recovering the true subclones, which cannot possibly
be represented as the assumed nonoverlapping clusters.

Finally, we carried out another simulation to
the sensitivity of the proposed inference to
different assumptions on experimental noise. In particular, we considered
experimental noise that
varies across SNVs, as it could arise from potential bias or
errors in data preprocessing, including sequencing bias, mapping bias,
errors in variant calling etc. Details of the simulation study are
reported in the supplementary material [\citet{supp}]. Briefly summarized,
in the simulation truth we replaced the error term $\eps_{t0}$ in
\eqref{eq:model_p} by an SNV-specific term $\varepsilon_{ts}$.
But we continued to fit the model with the common $\varepsilon_{t0}$,
as in
\eqref{eq:model_p}. We still find reasonable posterior inference.

For details, refer to the supplemental material.

\section{Pancreatic cancer data}
\label{sec:Example}
We analyzed NGS data obtained from exome sequencing of five samples of
pancreatic ductal
adenocarcinoma (PDAC) patients at NorthShore hospital. PDAC is a
particularly aggressive tumor with
median survival of less than a year. 
We extracted genomic DNA from each tissue and constructed an exome
library from these DNA using Agilent SureSelect capture probes. The
exome library was then sequenced in paired-end fashion on an Illumina
HiSeq 2000 platform. About 60 million
reads---each 100 bases long---were
obtained. Since the SureSelect exome was about 50 Mega bases, raw (pre-mapping)
coverage was about 120-fold. We then mapped the reads to the human
genome (version HG19)
[\citet{HG19}] using BWA [\citet{BWA}] and called variants using GATK
[\citet{GATK}]. Post-mapping, the mean coverage of the samples was
between 60 and 70 fold.

A total of nearly 115,000 SNVs and small indels were called
within the exome coordinates. We restricted our
attention to SNVs (i) that occur within genes that are
annotated to be related
to PDAC in the KEGG pathways database [\citet{KEGG}], (ii) that make a
difference to the protein translated from the gene, and (iii) that
exhibit significant coverage in all samples. This filtering left
us with $S=118$ SNVs.

\begin{figure}
\centering
\begin{tabular}{@{}cc@{}}

\includegraphics{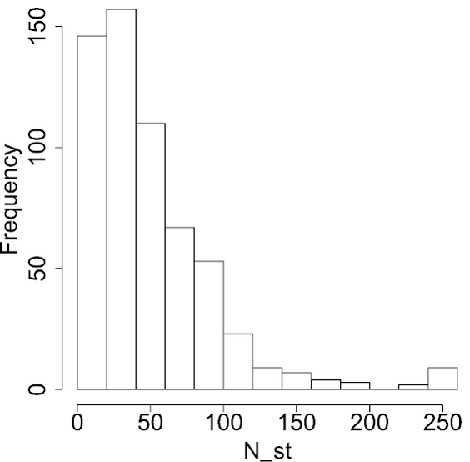}  & \includegraphics{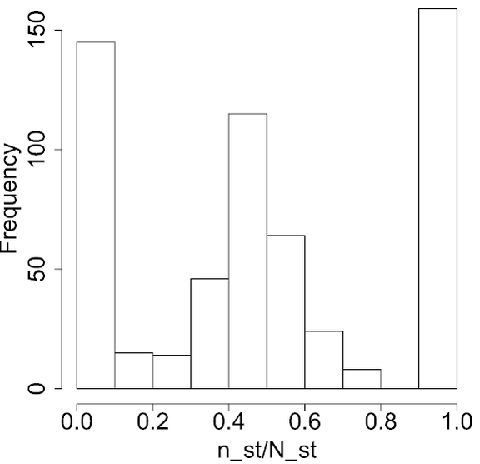}\\
\footnotesize{(a) $N_{st}$} & \footnotesize{(b) $n_{st}/N_{st}$}
\end{tabular}
\caption{Pancreatic cancer data: The left panel shows a histogram of the
total number of mapped reads, $N_{st}$, and the right panel shows a
histogram of the empirical fractions, $n_{st}/N_{st}$.}
\label{fig:data}
\end{figure}

\begin{figure}
\centering
\begin{tabular}{@{}cc@{}}

\includegraphics{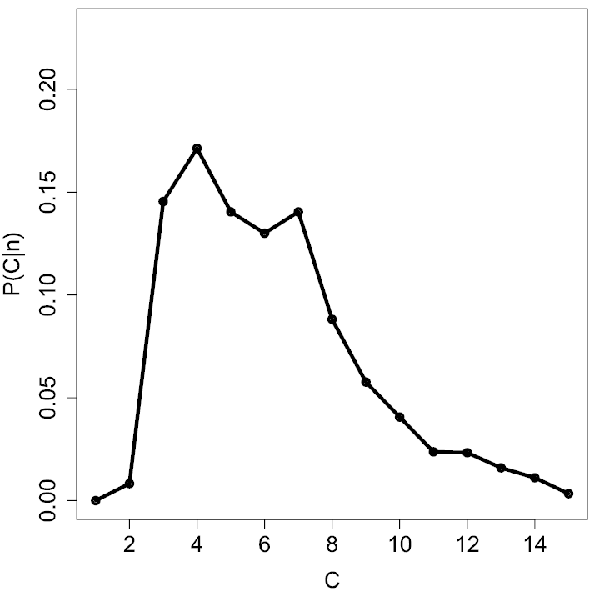}  & \includegraphics{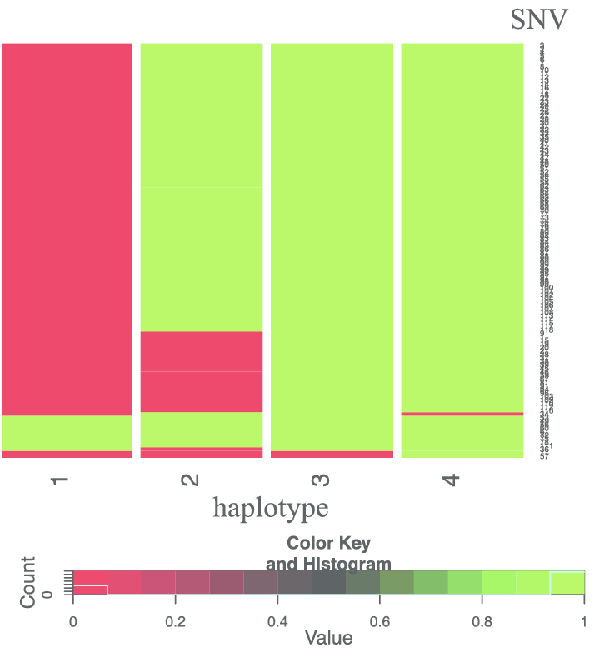}\\
\footnotesize{(a) $p(C \mid \mathrm{data})$} & \footnotesize{(b) $\bZ^\star$
with $C^\star=4$}
\end{tabular}
\vspace*{3pt}
\centering
\begin{tabular}{@{}c@{}}

\includegraphics{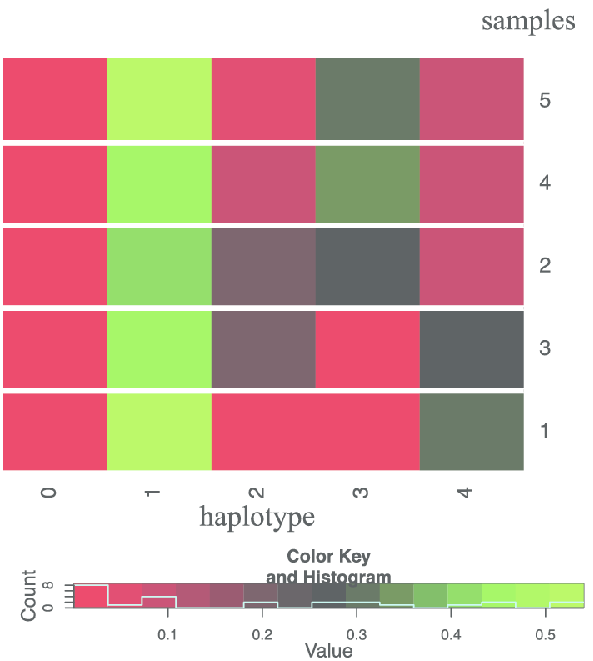}
\\
\footnotesize{(c) $\bw^\star$ with $C^\star=4$ and $\bZ^\star$}
\end{tabular}
\caption{Pancreatic cancer data: The posterior distribution of $C$ in
\textup{(a)}, the heatmaps of $\bZ^\star$ and
$\bw^\star$ with $C^\star=4$ in \textup{(b)} and \textup{(c)}, respectively. Note that
for $c=0$, $p^\star_0z^\star_{t0}$ is
illustrated in the first column of panel \textup{(c)}.}
\label{fig:panc_heatmap}
\end{figure}

In summary, using the earlier introduced notation, the data record
the read counts ($N_{st}$) and mutant allele read counts ($n_{st}$) of
$S=118$ SNVs from $T=5$ tumor samples. Figure~\ref{fig:data} shows
a summary of the data. The large $N_{st}$ values make the binomial
likelihood very informative.
%
%
For the prior specification, we let $r=0.2$, $\alpha=1$, $a=1$,
$a_0=1$, $a_{00}=5$ and $b_{00}=95$.
We generated $b_{st}$ from $\Be(25, 975)$ for the minimal training set.
We ran a MCMC
posterior simulation over
35,000 iterations, discarding an initial transient of 10,000
iterations. Figure~\ref{fig:panc_heatmap}(a) shows the marginal
posterior distribution for $C$. The posterior mode is $C^\star=4$.

%
%

The posterior point estimate of $\bZ$ conditional on $C^\star$ is
shown in Figure~\ref{fig:panc_heatmap}(b) and the corresponding
posterior point estimate of $\bw$ in
Figure~\ref{fig:panc_heatmap}(c).
Here, green represents a variant
sequence and red represents a reference sequence.
We find that each sample has two or three two dominant haplotypes, that is,
two green columns for each row in the heatmap.
Haplotypes 2, 3, 4 are shared by different sets of the five samples.
For example, sample 2
has a large-scaled abundance level for haplotypes 1, 2 and 3.
Sample 4 is mainly dominated by haplotypes 1 and 3.

These results indicate that while tumors are unique, there are
haplotypes that do recur across different patients. 
The results also clearly show that each tumor (in this data) is made of
more than one haplotype: usually two or three dominant haplotypes and
other minor types. 
To our knowledge, this is the first attempt to analyze the internal
clonal composition of multiple PDAC tumor samples based on NGS data.

%
\begin{figure}
\centering
\begin{tabular}{@{}cc@{}}

\includegraphics{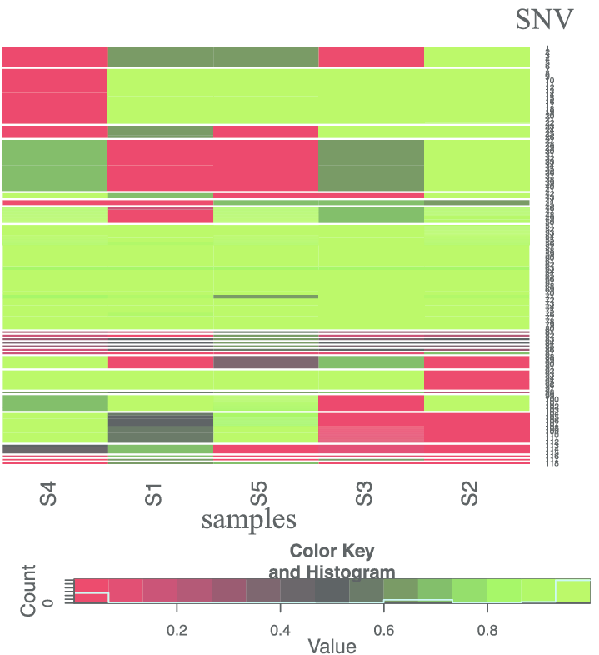}  & \includegraphics{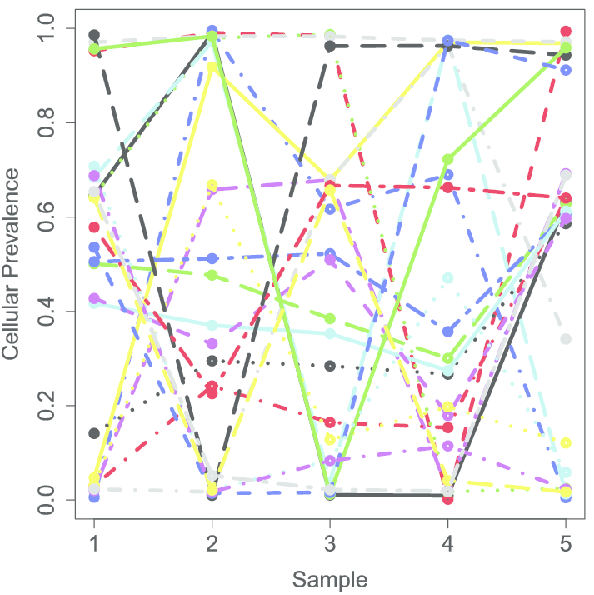}\\
\footnotesize{(a) Heatmap of cellular prevalence} & \footnotesize{(b)
Mean prevalence for each cluster}
\end{tabular}
\caption{Pancreatic cancer data: Estimated cellular prevalence of SNV
clusters over samples by PyClone.}
\label{fig:panc_pyclone}
\end{figure}
%

For comparison, we also evaluated tumor heterogeneity for the
same pancreatic cancer data using PyClone [\citet{roth2014pyclone}].
The results are shown in Figure~\ref{fig:panc_pyclone}.
The posterior estimated clustering includes 24 SNV clusters, shown in
panel (a). The estimated mean cellular prevalences of each cluster
across the five samples are shown in (b).
The estimated mean cellular prevalences vary substantially across samples.

\section{Lung cancer data}
\label{sec:lung_cancer}
We record whole-exome sequencing for four surgically
dissected tumor samples taken from the same patient with lung cancer.
The same bioinformatics preprocessing and analysis were
carried out as in\vadjust{\goodbreak} the previous pancreatic cancer example.
We obtained SNVs and filtered them based on criteria
similar to the previous example, leaving us in the end with
$S=101$ SNVs for the four intra-tumor samples.

\begin{figure}
\centering
\begin{tabular}{@{}cc@{}}

\includegraphics{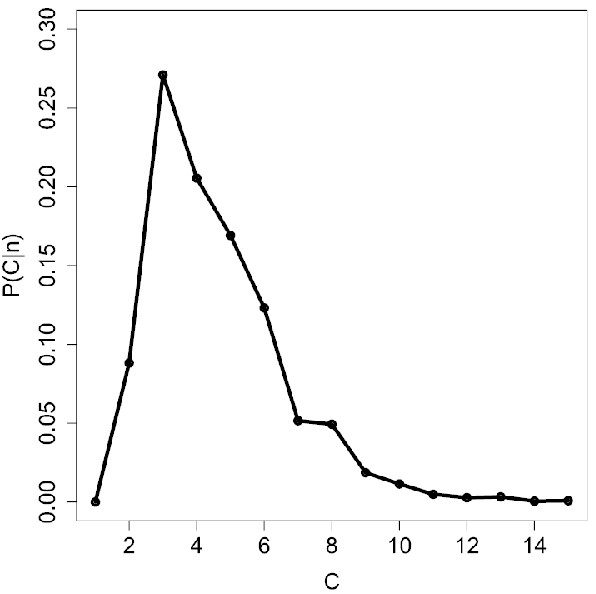}  & \includegraphics{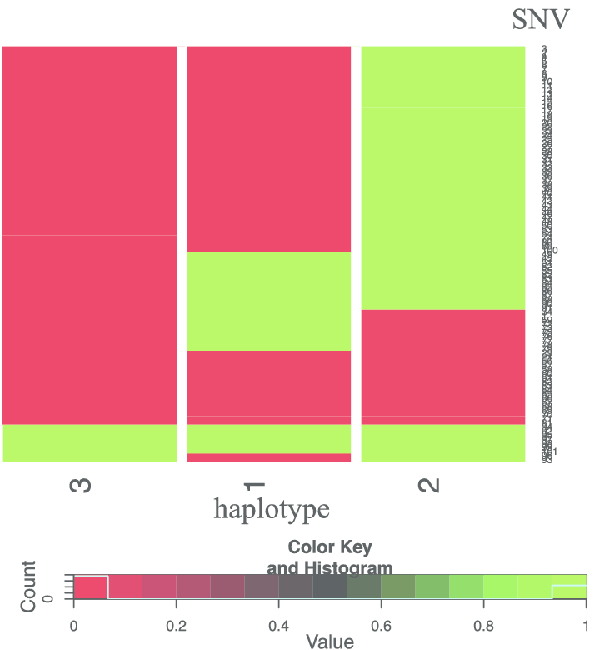}\\
\footnotesize{(a) $p(C\mid \mathrm{data})$} & \footnotesize{(b) $\bZ^\star$ with
$C^\star=3$}
\end{tabular}
\vspace*{3pt}
\centering
\begin{tabular}{@{}c@{}}

\includegraphics{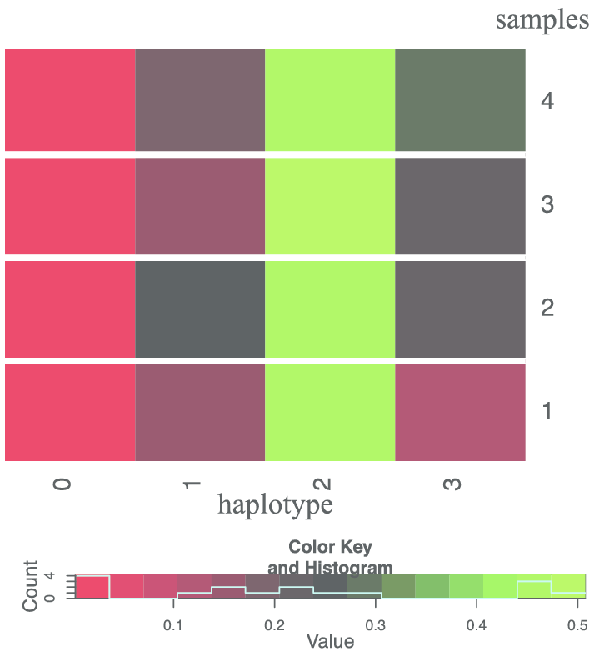}
\\
\footnotesize{(c) $\bw^\star$ with $C^\star=3$ and $\bZ^\star$}
\end{tabular}
\caption{Lung cancer data: The posterior distribution of $C$ in \textup{(a)},
the heatmaps of $\bZ^\star$ and $\bw^\star$ with $C^\star=3$ in
\textup{(b)} and
\textup{(c)}, respectively. Note that for $c=0$,
$p^\star_0z^\star_{t0}$ is illustrated in the first column of panel \textup{(c)}.}

\label{fig:lung_heatmap}
\end{figure}

We estimated the proposed Bayesian feature allocation model
with the same hyperparameters as in the previous PDAC data analysis.
Figure~\ref{fig:lung_heatmap}
summarizes the inference results. Panel (a) shows the marginal posterior
distribution for $C$, identifying a posterior mode at $C^\star=3$,
that is, three distinct haplotypes.
Panel (b) shows the posterior point estimate, $\bZ^\star_C$,
conditional on $\Cstar=4$.
The figure shows which SNVs are included for each of the three
haplotypes. Haplotype 3 contains
the smallest number of mutations (green bars), implying that
haplotype 3 might be the parental tumor cells. Haplotypes 1 and 2 are
descendants of haplotype 3 with additional somatic
mutations. Phylogenetically, a simple lineage can be hypothesized, with
haplotype 3 as the parent of haplotype 1 and/or
haplotype 2. 
Haplotype 2 possesses a large number of new
somatic mutations, potentially representing a type of aggressive tumor
cell. Panel (c) presents the posterior point estimate of $\bw$,
$\bw^\star$ with $C^\star$ and $\bZ^\star$.
Examining haplotypes 1--3, we found that, interestingly, all four tumor
samples share similar values of $\bw^\star_t$, implying a lack of
spatial heterogeneity across the tumor samples. In other words, these samples
all possess the inferred three tumor haplotypes in panel (b) with a
similar composition. 

%

\begin{figure}
\centering
\begin{tabular}{@{}cc@{}}

\includegraphics{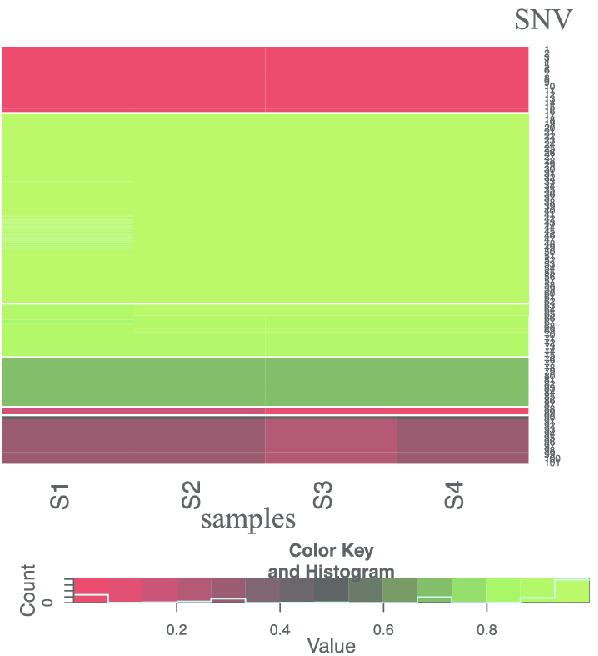}  & \includegraphics{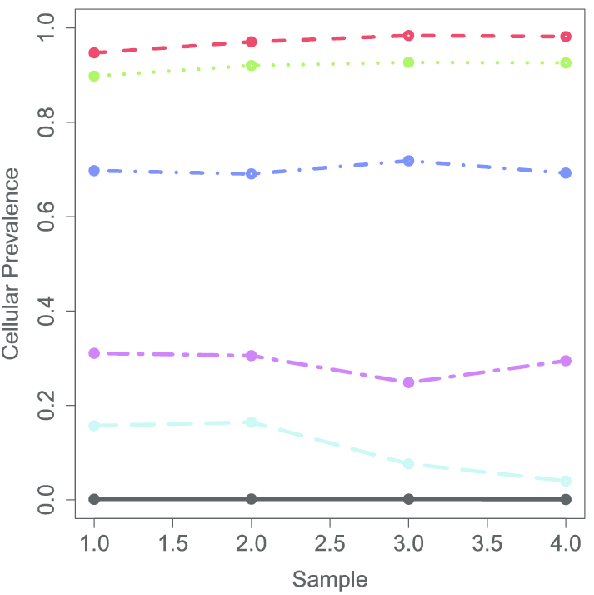}\\
\footnotesize{(a) Heatmap of cellular prevalence} & \footnotesize{(b)
Mean prevalence for each cluster}
\end{tabular}
\caption{Lung cancer data: Estimated cellular prevalence of SNV
clusters over samples by PyClone.}
\label{fig:lung_pyclone}
\end{figure}

Again, for comparison we also used PyClone [\citet{roth2014pyclone}]
with the lung cancer data.
The results are shown in Figure~\ref{fig:lung_pyclone}.
The estimated clustering identified six clusters of
mutations. The mean prevalences within a mutation cluster are similar
across samples.

%


\section{Conclusions}
\label{sec:Conclusion}

Tumors are heterogeneous tissues. The traditional way to identify
this heterogeneity has been to sequence multiple samples from the
tumor. Using such data to study the coexistence of genetically\vadjust{\goodbreak}
different subpopulations across tumors and within a tumor can shed
light on cancer development. Identifying subpopulations within a tumor
can lead to clinically important insights. For example,
\citet{CLL:2013:cell} found that a chemotherapy affects subclonal
heterogeneity in chronic lymphocytic leukemia. They also observed
that the presence of a certain subpopulation may adversely affect
clinical outcome.

We have proposed a model-based approach based on a feature
allocation model. The feature allocation model allows us to impute
inference about different components of tumor tissues based on NGS
data. 
The identified components are not necessarily unique
because there might be
other possible solutions which can lead to the same hypothetical
mutation frequencies.
Instead of reporting a single solution, the proposed approach provides
a full
probabilistic description of all possible solutions as a coherent posterior
probability model over $C$, $\bZ$ and $\bw$.

A number of extensions are possible for the present model. First, the
number of SNVs examined in this paper was relatively limited (about
100), although the total number of SNVs that were found in the whole
exome of
a tissue is on the order of about 50,000. Other than computational
complexity, there is no bar in principle on expanding the model to
analyze the whole SNV complement of the exome. It could also be
instructive to quantify the cellular diversity of the tumor based on
findings from various regions of the exome.

Another important extension of the model is in the basic
representation of subclones and haplotypes.
The current model uses a binary matrix to record whether
a variant sequence for an SNV is present or absent in a haplotype. A
variation of the model
could instead record for each subclone whether an SNV
is absent ($z_{sc}=0$), heterozygous ($z_{sc}=1$) or homozygous
($z_{sc}=2$). That is, $\bZ$ would become a trinary matrix.
Other extensions of the model are to
consider each SNV position to have four possible bases, ${A, C, G, T}$,
to introduce dependence among mutations or to formally model the noise
in variant calling. Each of these extensions is currently in
development. For example, incorporating
explicit error probabilities in variant calls is possible. Similar
to our previous work [\citet{ji2011bm}], we could replace the binomial
likelihood
\eqref{eq:like} in the
proposed model with a Bernoulli likelihood, for
each read, where the probability associated with a read depends on
quality scores of base calling and read mapping. We will consider
this extension as future work.


Tumor genome sequencing projects have typically looked for specific
genes to be mutated or not. The inherent assumption here, so far
unproven, is that the overall effect of carcinogenesis could be
explained by a handful of changes in a small number of genes. Our
model takes the opposite approach and allows us to examine the whole
genome (or exome) and, by considering VAF patterns, to construct
reasonable models for the tissue. We believe this holistic approach to
the analysis might provide more robust conclusions and biomarkers than
the gene-by-gene approach.


\begin{supplement}[id=suppA]
\stitle{Supplement to
``A Bayesian feature allocation model for tumor
heterogeneity''}
\slink[doi]{10.1214/15-AOAS817SUPP} 
\sdatatype{.pdf}
\sfilename{aoas817\_supp.pdf}
\sdescription{The
supplementary material includes the second simulation study.}
\end{supplement}




%
%




\printaddresses
\end{document}